 \documentclass[aps,prc,twocolumn,showpacs,showkeys,amsmath,amssymb,a4paper]{revtex4-1}
\usepackage{amssymb}
\usepackage{amsbsy}
\usepackage{graphics}
\usepackage{psfrag}
\usepackage{epsfig}
\usepackage{multirow,mathtools}
\usepackage{bm}
\usepackage{color}
\usepackage{soul}
\def\bea{\begin{eqnarray}}
\def\eea{\end{eqnarray}}
\def\beq{\begin{equation}}
\def\eeq{\end{equation}}

\newcommand{\Lc}{{\Lambda}_c}
\newcommand{\Sc}{{\Sigma}_c}

\begin{document}
\title{Comment on 
``$\Lambda_c N$ interaction in leading order covariant chiral effective field theory''
}
\author{J. Haidenbauer$^1$}
\email{j.haidenbauer@fz-juelich.de}
\affiliation{
$^1$Institute for Advanced Simulation, Institut f\"ur Kernphysik, and J\"ulich Center for Hadron
Physics, Forschungszentrum J\"ulich, D-52425 J\"ulich, Germany 
}
\author{G. Krein$^2$}%
\email{gastao.krein@unesp.br}
\affiliation{$^2$Instituto de F\'{\i}sica Te\'orica,
Universidade Estadual Paulista,
Rua Dr. Bento Teobaldo Ferraz, 271 - Bloco II,
01140-070 S\~ao Paulo, SP, Brazil}

\begin{abstract}
Song et al. [Phys. Rev. C 102, 065208 (2020)] presented results for the
$\Lambda_c N$ interaction based on an extrapolation of lattice simulations
by the HAL QCD Collaboration at unphysical quark masses to the physical
point via covariant chiral effective field theory. We point out that 
their predictions for the $^3D_1$ partial wave disagree with
available lattice results. We discuss the origin of that disagreement
and present a comparison
with predictions from conventional (non-relativistic) chiral effective field 
theory. 
\end{abstract}

%\pacs{13.60.Rj,14.40.Lq,25.43.+t}

\maketitle

Due to the lack of any experimental information, at present results from lattice 
QCD simulations provide the only model-independent estimate for the strength
of the interaction of charmed baryons with nucleons. 
Corresponding lattice calculations for the $\Lc N$ and $\Sc N$ systems have 
been published by the HAL QCD Collaboration 
\cite{Miyamoto:2017,Miyamoto:2018,Miyamoto:2019}, though for unphysical quark 
masses corresponding to pion masses of $m_\pi = 410 - 700$ MeV. 
Evidently, in order to draw conclusions on the physics implications and to allow 
for predictions for future experiments an extrapolation of the HAL QCD results 
to the physical point is
required. Such an extrapolation was performed by us in Ref.~\cite{Haidenbauer:2018}, 
for $\Lc N$ and in \cite{Haidenbauer:2020} for $\Sc N$. We used as guideline
conventional (non-relativistic) chiral effective field theory ($\chi$EFT) \cite{Epelbaum:2003} 
up to next-to-leading order (NLO).
With regard to the $\Sc N$ interaction, another extrapolation of the HAL QCD 
results was performed in Ref.~\cite{Meng:2020} using heavy baryon 
chiral perturbation theory and taking into account heavy quark spin symmetry.

In this comment we critically discuss results for the $\Lc N$ interaction 
of yet another and very recent extrapolation done by Song et al. \cite{Song:2020} 
which utilizes covariant $\chi$EFT. The corresponding work is based on a potential 
at leading-order (LO) in the chiral expansion. 
Song et al. \cite{Song:2020} found that in the case of the 
$^1S_0$ partial wave the lattice QCD data at 
unphysical masses ($m_\pi = 410,\, 570$~MeV) can be quite well reproduced within
covariant $\chi$EFT. At the physical point, the predicted phase shifts are rather 
similar to those we obtained in Ref.~\cite{Haidenbauer:2018}. 

For the coupled $^3S_1$-$^3D_1$ partial waves the situation is quite different. Here 
the $^3S_1$ phase 
shifts from covariant $\chi$EFT are only in fair agreement with the lattice QCD data,
see Fig. 4 (left side) in Ref.~\cite{Song:2020}. Specifically for $m_\pi = 570$~MeV
the energy dependence suggested by the HAL QCD results is not properly reproduced. 
Moreover, the extrapolation to the physical point yields rather different results 
as compared to the predictions in Ref.~\cite{Haidenbauer:2018}. 
While conventional $\chi$EFT suggests a
moderately attractive $^3S_1$ interaction, with phase shifts almost identical to the 
one in the $^1S_0$ partial wave, the covariant approach leads to a predominantly 
repulsive result which, in addition, exhibits a strong energy dependence. 
 
\begin{figure*}[t]
 \begin{center}
\includegraphics[height=85mm,angle=-90]{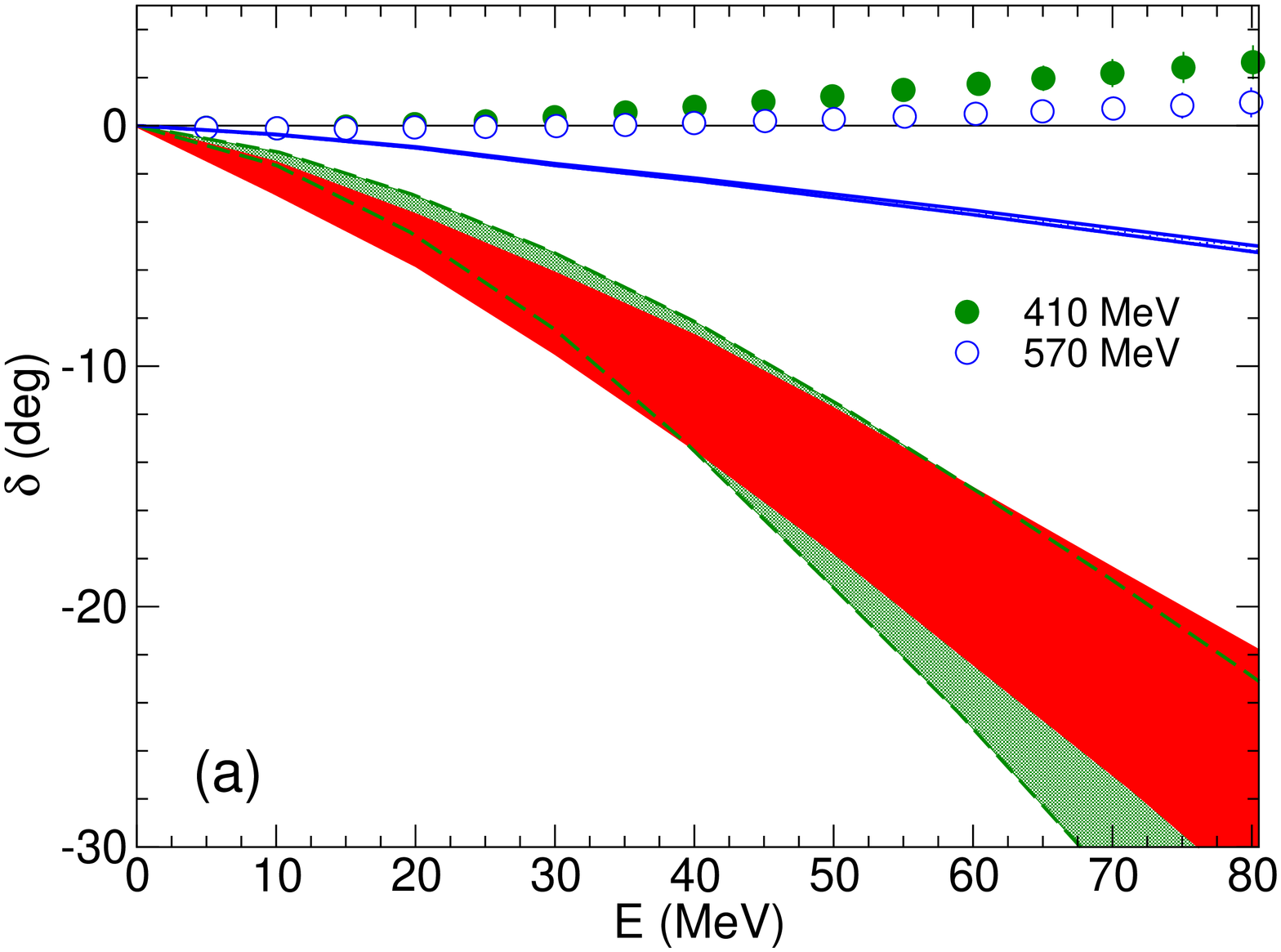}
\includegraphics[height=85mm,angle=-90]{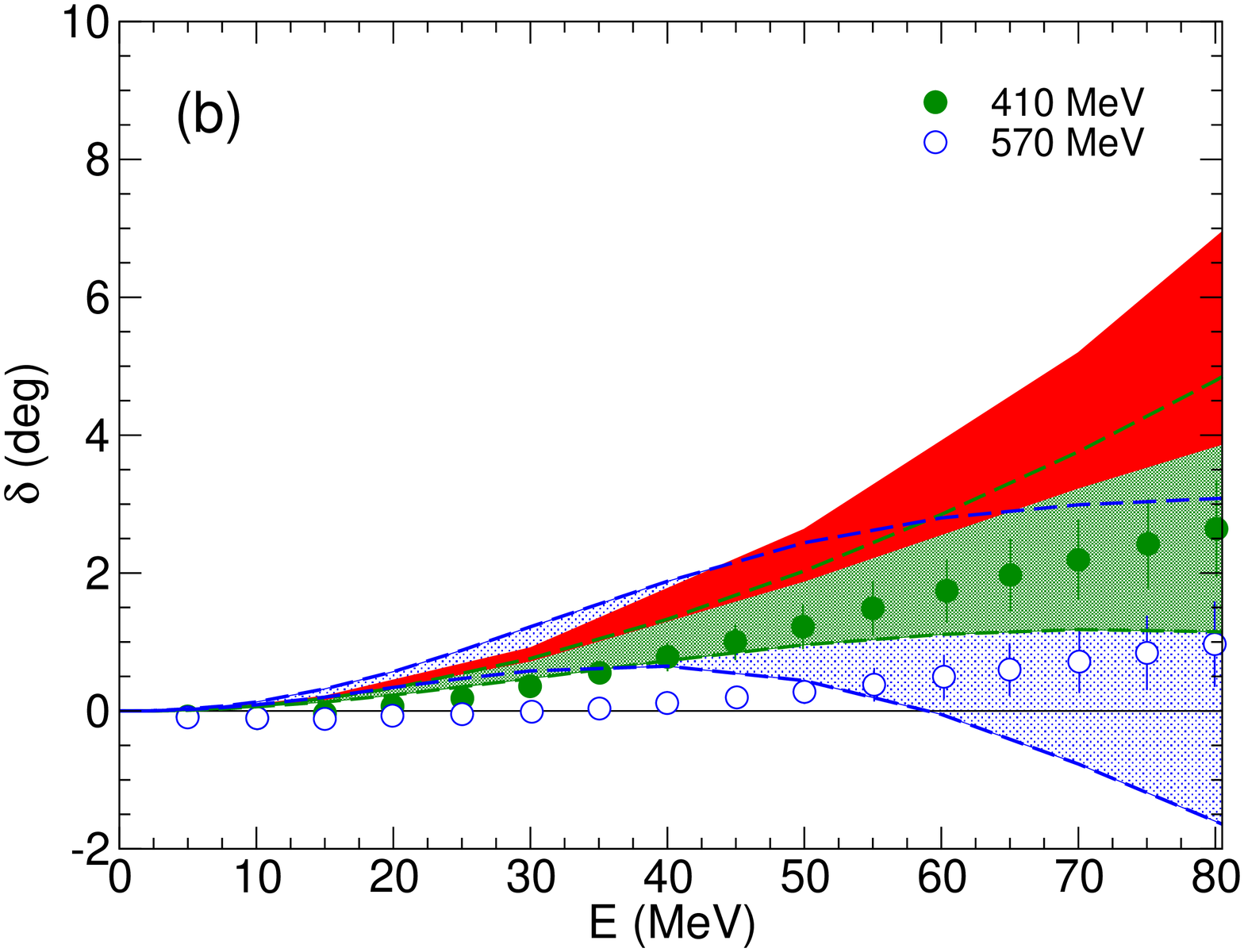}
\caption{Predictions for the $\Lc N$ $^3D_1$ phase shift.  
(a) Results from covariant $\chi$EFT taken from Ref.~\cite{Song:2020}. 
(b) Results based on the $\Lc N$ potential from Ref.~\cite{Haidenbauer:2018}.
 Red (black), green (dark grey), 
and blue (light grey) bands correspond to 
$m_\pi=138$, $410$, and $570$~MeV, respectively.
The width of the bands represent cutoff variations/uncertainties. 
Lattice results of the HAL QCD Collaboration corresponding to
$m_\pi=410$~MeV (filled circles) and $570$ MeV (open circles) 
are taken from Ref.~\cite{Miyamoto:2019}. Note the different scales of (a) and (b)! 
}
\label{P3d1}
 \end{center}
\vskip -0.5cm
\end{figure*}

As a byproduct of their study, Song et al. provided also 
predictions for the $^3D_1$ phase shifts and the mixing angle $\varepsilon_1$ --
at the physical point and for the pion masses of the HAL QCD simulation --
with the intention that those can be checked by future lattice QCD calculation. 
Interestingly, such results are already available. They have been presented in 
the PhD thesis by Takaya Miyamoto~\cite{Miyamoto:2019} which can be accessed via
Inspire. 
We show the lattice results for $^3D_1$ in Fig.~\ref{P3d1}(a) together with 
the predictions from covariant $\chi$EFT. Obviously, there is a strong mismatch. While 
lattice QCD suggests a weakly attractive interaction, the results by Song et al. are 
strongly repulsive. Corresponding results of our $\Lc N$ interaction 
\cite{Haidenbauer:2018}, which are likewise predictions, are displayed in 
Fig.~\ref{P3d1}(b). Evidently, in case of conventional $\chi$EFT there is a 
remarkable qualitative agreement with the lattice simulation.
Specifically, the results for $m_\pi = 410$~MeV are rather well in line with those
by the HAL QCD Collaboration. This gives us confidence that our extrapolation 
to the physical point is plausible and reasonable, for $^3D_1$  as well as for
$^3S_1$.
%As emphasized in Ref.~\cite{Haidenbauer:2020}, at the NLO level the reliability of 
%the results is limited to energies up to roughly $50$~MeV. 

For understanding the origin of the difference let us discuss briefly the main
features of the two approaches. 
For further details, specifically how the pion-mass dependence is implemented,  
we refer the reader to Ref.~\cite{Song:2020} with regard to the
extrapolation based on covariant $\chi$EFT and to Ref.~\cite{Haidenbauer:2018}
for the conventional approach. In either
frameworks the potential is given in terms of pion exchanges and a series of contact 
interactions with an increasing number of derivatives. The latter represent the short-range
part of the baryon-baryon force and are parameterized by low-energy constants (LECs), that
need to be fixed by a fit to data \cite{Epelbaum:2008}. 
The essential difference in the corresponding potentials
occurs in the contact terms and arises from the circumstance that in the covariant formulation
the potential is derived with the full Dirac spinor of the baryons included \cite{Song:2020}.  
As a consequence, while in conventional $\chi$EFT based on the Weinberg 
counting~\cite{Epelbaum:2008}, contributions to the contact interaction 
of chiral power $\nu$ are proportional to $p^\nu$ (with $p$ being the modulus of the 
baryon-baryon center-of-mass momentum) and, thus, are uniquely related to the order of the 
chiral expansion, this no longer
the case for the covariant version. Moreover, in the conventional $\chi$EFT 
a partial wave expansion of the contact interaction with the involved spin- and momentum-dependent 
operators allows one to rewrite the various contributions in terms of suitably defined LECs 
that then contribute only to single partial waves. 
This is not possible in the covariant power counting, in which already the LO contact 
term contributes to all $J=0,1$ partial waves and the potential strengths
in the various partial waves, characterized by the LECs, are interrelated. 
That aspect impacts the extrapolated results in Ref.~\cite{Song:2020}. 
Specifically, there is a contribution to the $^3D_1$ 
potential from the contact interactions already at LO where the corresponding 
LECs are those that appear likewise in the $^1S_0$ and $^3S_1$ partial waves, 
cf. Eq.~(7) in Ref.~\cite{Song:2020}. Apparently, fixing those LECs solely from the 
$S$-wave phase shifts of the lattice simulations leads to wrong results for the $^3D_1$.
In the Weinberg counting a $^3D_1$ contact interaction arises first at 
next-to-next-to-next-to-leading order (N$^3$LO) \cite{Epelbaum:2008} and 
it is independent of the LECs in the $S$-waves! 

In summary, conventional $\chi$EFT (up to NLO) employed in Ref.~\cite{Haidenbauer:2018}
seems to provide a more reliable tool for representing lattice QCD results by the HAL QCD
Collaboration for the $\Lc N$ interaction at $m_\pi = 410,\,570$~MeV \cite{Miyamoto:2019}
and for extrapolating them to the physical point.
In covariant $\chi$EFT as utilized by Song et al. \cite{Song:2020}
(at LO) it is obviously more difficult to account for the results at unphysical pion
masses on a quantitative level, specifically for the spin triplet
case, i.e. in situations where coupled-channel effects could be important.
As a consequence, extrapolations are more unstable.
In any case, lattice simulations for quark masses closer to the physical point would
be rather useful to shed further light on the issue of extrapolation
and, of course, direct experimental constraints \cite{Haidenbauer:2020kwo} would be helpful too.

%%%%%%%%%%%%%%%%%%%%%%%%%%%%%%%%%%%%%%%%%%%%%%%%%%%%%%%%%%%%%%%%%%%%%%%%%%%%%%%%

\end{document}